\documentclass[twocolumn,showpacs,preprintnumbers,amsmath,amssymb]{revtex4}

\usepackage{graphicx}
\usepackage{dcolumn}
\usepackage{bm}

\begin{document}

\newcommand{\be}{\begin{equation}}
\newcommand{\ee}{\end{equation}}
\newcommand{\bea}{\begin{eqnarray}}
\newcommand{\eea}{\end{eqnarray}}
\newcommand{\non}{\nonumber}


\title{A momentum-space representation  of  Green's functions \\ 
   with modified dispersion  on ultra-static space-time.}
\author{Massimiliano Rinaldi}
\email{rinaldim@bo.infn.it}

\affiliation{ Dipartimento di Fisica, Universit\`a di Bologna and 
 I.N.F.N. Sezione di Bologna,
V. Irnerio 46, 40126 Bologna, Italy.\\
Museo Storico della Fisica e Centro Studi e Ricerche ``E. Fermi",\\ Compendio Viminale, 00184 Rome, Italy.} 
\date{\today}
\begin{abstract}
We consider the Green's functions associated to a scalar field propagating on a curved, ultra-static background, in the presence of modified dispersion relations. The usual proper-time deWitt-Schwinger procedure to obtain a series representation of the Green's functions is doomed to failure, because of higher order spatial derivatives in the Klein-Gordon operator. We show how to overcome this difficulty by considering
a preferred frame, associated to a unit time-like vector. With respect to this frame, we can express the Green's functions as an integral over all frequencies of a space-dependent function. The latter can be expanded in momentum space, as a series with geometric coefficients similar to the deWitt-Schwinger's ones. By integrating over all frequencies, we finally find the expansion of the Green's function up to four derivatives of the metric tensor. The relation with the proper-time formalism is also discussed.
\end{abstract}


\maketitle

\section{Introduction}

Modified dispersion relations (MDRs) have recently  attracted a flurry activity in various  high-energy physics  models. The common motivation is that MDRs can be used as a phenomenological approach to investigate physics at the Planck scale, where General Relativity is no longer reliable. In general, the lack of a complete theory of quantum gravity, leads to consider trans-Planckian effects as perturbative above a certain energy scale. The first application is certainly in cosmology. Several authors believe that trans-Planckian effects affected the early-stage evolution of the Universe, and that they left some observable fingerprints, e.g. in the CMB inhomogeneities (see, for example, \cite{MDRcosmo}). A similar situation occurs in black hole physics, where the quantum thermal emission discovered by Hawking \cite{HAW} is related to modes of arbitrarily large frequency near the horizon. In this case, it has been proved that the spectrum emitted at infinite distance from the hole is only marginally affected by MDRs \cite{Unruh1}. 

Some fundamental theories, such as String Theory, Loop Quantum Gravity, and Double Special Relativity,  predict MDRs (see the review \cite{mattingly} for references). 
However, little is known about the modifications that MDRs generate in the formalism of quantum field theory on curved space. For example, the renormalization of the stress tensor is crucial to evaluate the back-reaction in the semi-classical theory \cite{BirDav}. Such a quantity, in the presence of MDRs, has been recently obtained in the context of cosmology, by the authors of \cite{Mazzitelli2}.  In this work, it is shown that the renormalization procedure leads to a rescaling of the bare Newton's constant and cosmological constant. However, in the case of non-homogeneous backgrounds, the renormalization of the stress tensor, in the presence of MDRs, appears much more difficult, as the Klein-Gordon operator, unlike the cosmological case, now contains spatial derivatives of (at least) fourth order  (see for example \cite{Corley}). In fact, the usual deWitt-Schwinger representation of the Green's functions, which is the starting point of the point-splitting technique, \cite{dewitt}-\cite{fulling} does not work in this case.

In this paper, we begin to consider the problem of MDRs on non-homogeneous manifolds by first looking at the ultra-static case. Our method, which should be applicable also to stationary metrics, relies upon the existence of a preferred frame. It is known that, even though MDRs break the local Lorentz invariance,  general covariance can be preserved by introducing a preferred frame through a dynamical unit time-like vector field $u^{\mu}$ \cite{Jacobson1}-\cite{LLMU}. With the help of the latter, we can foliate the manifold into space-like surfaces. Furthermore, if the space-time is also ultra-static, the Green's functions can be written as an integral over all frequencies of a function, which is independent of the time associated to the observer co-moving with $u^{\mu}$. This function satisfies an equation, which can be solved in momentum space by applying the well-known Bunch and Parker procedure \cite{BP}. Thus, we can find a momentum-space representation of the time-independent part of the Green's functions.
 
The plan of this work is the following: in the next section, we introduce the modified dispersion relation which will be used in this work, and, in section 3, we calculate the two-point function in flat space. In section 4 we introduce the deWitt-Schwinger analysis, suitably adapted to our case. In section 5 we present the Bunch and Parker method, and we find the expansion of the Green's function in momentum space up to four derivatives of the metric. Finally, we conclude with some remarks and further conjectures. Throughout this paper, we use the signature $(-,+,+,+\cdots)$, and set $\hbar=c=1$

\section{Modified dispersion relations}

\noindent In Minkowski space-time the dispersion
relation for a  scalar field of mass $m$ can be found by inserting the ansatz $\phi \sim
\exp(-ik_0t+i\vec k\cdot \vec x)$ in the Klein-Gordon
equation \bea (-\nabla^2+m^2)\phi=0~,\eea which, together with the identity $\omega_k=|k_0|$, yields
\bea \omega_k^2=|\vec k|^2+m^2~.\eea A general
dispersion relation can be written as \bea \omega_k^2=|\vec k|^2+{\cal F}(|\vec k|)+m^2~, \eea where ${\cal F}(|\vec k|)$ is a scalar
function of the modulus of the wave-vector $\vec k$. If ${\cal F}$ depends on the square of the modulus, rotation invariance is preserved. If it is also
analytic, then it can be expanded, and, to leading order, the MDR reads \bea\label{MDR} \omega_k^2=|\vec k|^2+\epsilon^2
|\vec k|^4+m^2~, \eea where $\epsilon$ is a cut-off parameter that
sets the lowest value of $\vec k$ at which corrections to the
dispersion relation are ignored. Also, the sign of $\epsilon^2$
indicates wether the modes are sub-luminal ($\epsilon^2 <0$) or
super-luminal ($\epsilon^2>0$). These kinds of MDRs  were considered in Cosmology, but also in the context of the
analogue models of gravity constructed with superfluids \cite{Unruh1}. 
It is clear that MDRs, such as (\ref{MDR}),  break Lorentz invariance. However, general covariance
can be preserved if the preferred frame is associated to a dynamical
quantity. This is the route followed by Jacobson et al.
\cite{Jacobson1}: the preferred frame is determined
by a unit timelike vector field $u^a$, which enters quadratically
the action. The latter has the form
\begin{equation}
S=\int
d^4x\sqrt{-g}\left(-a_1R-b_1F^{ab}F_{ab}+\lambda(g_{ab}u^au^b+1)\right)~,
\end{equation}
where $a_1$ and $b_1$ are constant parameter,
$F_{ab}=2\nabla_{[a}u_{b]}$, and $\lambda$ is a Lagrange multiplier
that ensures $u^a$ to be a unit time-like vector. It is then possible to construct a massless scalar field Lagrangian, which preserves general covariance, and reads
\bea\label{scalaraction} {\cal L}_{\varphi}=\frac{1}{2}\left((\nabla
\varphi)^2+\epsilon^2(\hat\nabla^2\varphi)^2\right)~,\eea where $\epsilon$ sets the scale at which Lorentz invariance breaks. The operator $\hat\nabla^2$ is the
covariant spatial Laplacian defined as \bea\label{spacelap}
\hat\nabla^2\varphi=-q^{ac}\nabla_{a}(q_c^{\,\,\,b}\nabla_b\varphi)~,\eea where $q_{ab}$ is the induced metric on the spatial sections, defined
by \bea q_{ab}=g_{ab}+u_au_b~.\eea

The inclusion of the vector field $u^a$ as a dynamical variable
leads to a theory similar to the Maxwell-Einstein action. Therefore,
one expects an Abelian conserved charge associated to $u^a$. However, it is possible to show that if one assumes a FLRW flat
metric, the equations of motion are trivially solved, i.e.
$F_{ab}=0$. On the contrary, symmetries do not allow to choose $F_{ab}=0$, together with a Schwarzschild metric. In fact, the
conserved charge associated to $F_{ab}$ does not vanish, leading to a Reissner-Nordstr\"{o}m form for the metric
\cite{Jacobson1}. Therefore, when MDRs associated to dynamical
vector fields are considered, the modes propagate on
a modified metric. In this case, one can argue that the
modifications introduced by $u^a$ can be somehow set to be very
small, so that the modes with a MDR can still be
seen as propagating on an unchanged background. In any case, in order to evaluate the effects of MDRs at the level of back-reaction, we first need to compute the renormalized Green's functions. In order to do so, we begin by evaluating the deWitt-Schwinger expansion in section 4 and 5, while in the next section we look at the simple flat space case.

\section{MDRs in flat space}

Before embarking upon the study of the curved background case, let us have a look first to what happens in flat space. For simplicity, we consider the 2-dimensional Minkowski metric
\bea ds^2=-dt^2+dx^2~,
\eea where the Klein-Gordon equation, obtained from the Lagrangian (\ref{scalaraction}), reads
\bea\label{2Dflat}
\nabla^2\phi(x,t)-\epsilon^2\partial_x^4\phi(x,t)=0~,\quad \epsilon^2>0~.
\eea
Despite the modification, the standard function $\exp (-i\omega t+ipx)$ is a basis for the (physical)  solution space, provided $\omega_p^2=p^2+\epsilon^2p^4$. This might be confusing, since we expect \emph{four} independent solutions. Indeed, this is the case but two out of the four solutions have \emph{imaginary} momenta. To show this, it is sufficient to replace the ansatz $\phi(x,t)\sim\varphi(x)\exp (-i\omega_p t)$ into Eq.~(\ref{2Dflat}), and find the fourth-order differential equation
\bea
\epsilon^2\partial_x^4\varphi(x)-\partial_x^2\varphi(x)-\omega^2\varphi(x)=0~.
\eea
The latter has four independent solutions of the form
\bea
\varphi_{1,2}(x)=C_{1,2}\,e^{\pm\Omega^{(-)}x}~,\quad \varphi_{3,4}(x)=C_{3,4}\,e^{\pm\Omega^{(+)}x}~,
\eea
where $C_{1,2,3,4}$ are normalization constants, and
\bea
\Omega^{(\pm)}=\frac{\sqrt{1\pm\sqrt{1+4\epsilon^2\omega_p^2}}}{\sqrt{2}\epsilon}~.
\eea
We note that $\Omega^{(-)}$ is a pure imaginary number and that, for $\epsilon\rightarrow 0$, it tends to the usual factor $i\omega_p$. On the contrary, $\Omega^{(+)}$ is real and it diverges when $\epsilon\rightarrow 0$. By solving the above identity for $\omega_p$, one finds that
\bea
\omega_p^2=\pm(\Omega^{(\pm)})^2+\epsilon^2(\Omega^{(\pm)})^4~,
\eea
which is nothing but the modified dispersion relation, provided we identify $\Omega^{(-)}$ with $ip$. It then follows that $\Omega^{(+)}=p$ and the corresponding functions $\varphi_{3,4}$ span an unphysical sector of the solution space \footnote{In the case of subluminal dipersion, all four roots can be real for $\omega_p$ small enough.}.

In momentum space, the massless propagator corresponding to the modified dispersion reads
\bea
G(p)=\frac{1}{p_{\mu}p^{\mu}+\epsilon^2 p^4}=\frac{1}{\omega_p^2-p_0^2}~.
\eea
To compute the two-points function, it is convenient to find first the Wightman functions $G^+$ and $G^-$, by choosing the appropriate contour ${\cal C}$ in the 2-dimensional Fourier transform\footnote{In two dimensions, the vector product $\vec{p}\cdot\vec{x}$ has to be considered as a sum over left-moving ($-x$) and right-moving ($+x$)  waves.}
\bea
G(x^{\mu},x'^{\mu})=\int_{\cal C} \frac{dp_0dp}{(2 \pi)^2}\frac{e^{ip^{\mu}\Delta x_{\mu}}}{\omega_p^2-p_0^2}~,
\eea
where  $\Delta x^{\mu}=x^{\mu}-x'^{\mu}$. The pole structure of the above integral is formally the same as in the relativistic case: by adding a small imaginary part to the denominator, the poles are located at $p_0=\pm \omega_p\mp i\delta$, with $\delta<<1$.  The functions $iG^+$ and $iG^-$ can be found directly by choosing ${\cal C}$ as circles around each pole (clockwise around $-\omega_p$ and counter-clockwise around $+\omega_p$, see \cite{BirDav}). The result is
\begin{equation} G^{\pm}(x^{\mu},x'^{\mu})={1\over 2\pi}\int _{\Lambda}^{+\infty}\frac{\cos(p\Delta x )}{ \sqrt{p^2+\epsilon^2 p^4}}\,e^{\left(\mp i\Delta t\sqrt{p^2+\epsilon^2 p^4}\right)}dp~,\label{wightman}
\end{equation}
where $\Lambda$ is the usual IR cut-off, and $\Delta x=x'-x$, $\Delta t=t'-t$. The breaking of Lorentz invariance is now manifest, since $\Delta x$ and $\Delta t$ do not have the same $p$ pre-factor. This integral cannot be solved exactly, hence some approximations need to be done. First, we note that the ratio between the integrand in Eq.~(\ref{wightman}) and the integrand of the relativistic propagator 
\bea 
G^{\pm}_{\rm{rel}}(x^{\mu},x'^{\mu})={1\over 2\pi}\int _{\Lambda}^{+\infty}\frac{\cos(p\Delta x)}{p}\,e^{(\mp ip\Delta t )}dp~,
\eea
tends to one as $p\rightarrow 0^+$. It is then reasonable to assume that the IR behavior of the modified Green's function is the same as the relativistic one. Therefore, we do not worry too much about the IR divergence and we look at the large $p$ regime, by approximating the integral (\ref{wightman}) as 
\begin{equation}
\frac{d}{d\Delta t}(G^++G^-)=-{1\over 2\pi}\int_0^{+\infty}dp\cos (p\Delta x)\sin(\epsilon p^2\Delta t)~,
\end{equation}
which yields
\begin{widetext}
\bea\label{flatG}
\langle \phi^2(x^{\mu},x'^{\mu}) \rangle =G^++G^-=-\sqrt{\frac{\Delta t}{2\epsilon \pi}}\left[\cos\left(\frac{\Delta x^2}{4\epsilon\Delta t}\right)-\sin\left(\frac{\Delta x^2}{4\epsilon\Delta t }\right)\right]-\frac{\Delta x}{2\epsilon}\left[C\left(\frac{\Delta x}{\sqrt{2\pi \epsilon\Delta t}}\right)+S\left(\frac{\Delta x}{\sqrt{2\pi \epsilon\Delta t}}\right)\right]~,
\eea
\end{widetext}
where $C$ and $S$ are Frenel's integrals\footnote{These are defined as  $S(x)=\pi/2\int_0^{x}\sin t^2dt$ and  $F(x)=\pi/2\int_0^{x}\cos t^2dt$.}. It is interesting to note that in the coincidence limit $x^{\mu}=x'^{\mu}$, this expression is actually \emph{finite}. However, we expect divergences in the coincidence limit of the energy-momentum tensor, as the latter is calculated trough derivatives of the two-point function with respect to $\Delta t$ and $\Delta x$.

\section{The deWitt-Schwinger expansion}

In the previous Section we found an expression for the two-point function at large momenta and in flat space. In the relativistic case, the generalization of such expressions on curved space proceeds through the deWitt-Schwinger construction \cite{dewitt}-\cite{fulling}, which works directly in coordinate space.
Here we briefly outline the procedure, adapted to the modified dispersion relation (\ref{MDR}), with $\epsilon^2>0$. We consider a $n+1$-dimensional  globally hyperbolic manifold, such that it can be foliated into space-like surfaces of constant $\tau$ \cite{LLMU}. The parameter $\tau$ can be used to define the unit time-like tangent vector $u_{\mu}=-\partial_{\mu}\tau$, with respect to some coordinate system $x^{\mu}$. Therefore, $\tau$ assumes the r\^ole of the time relative to a free-falling observer moving with velocity $u^{\mu}$. Thus, we can write the metric as \bea\label{Jmetric}
ds^2=g_{\mu\nu}dx^{\mu}dx^{\nu}=-d\tau^2+q_{\mu\nu}dx^{\mu}dx^{\nu}~.\eea where $d\tau=u_{\mu}dx^{\mu}$. In this paper, we consider the case when $q^{\mu\nu}=g^{\mu\nu}+u^{\mu}u^{\nu}$ does not depend on $\tau$, i.e. when the metric is ultra-static
The form of this metric guarantees that
 \bea\label{dets} 
 \det g_{\mu\nu}= \det q_{ij}~,
 \eea  
where Latin indices, here and in the following, label spatial coordinates only. Thus, the d'Alambertian operator acting on a scalar $\phi$ can be decomposed as
\bea\label{comder}
\nabla^2\phi=\hat\nabla^2\phi-\ddot\phi~,
\eea where the dot indicates derivative with respect to $\tau$.

Modified dispersion relations (\ref{MDR}) naturally appears if one considers the Klein-Gordon equation 
\bea\label{modklein}
(\nabla^2-m^2-\epsilon^2\hat\nabla^4)\,\phi(x)=0~,\eea for which the associate Green's functions satisfies the equation
\begin{equation}
(\nabla^2-m^2-\epsilon^2\hat\nabla^4)\,{\cal G}(x,x')=-g^{-1/2}\delta^{(n+1)}(x-x')~.
\end{equation} Here, and from now on, we denote $g(q)\equiv \det | g(q)|$. Because of the MDR, the function ${\cal G}$ is no longer Lorentz invariant, but, since the modification is a quartic spatial operator, the $O(n)$ rotational invariance holds. It is therefore convenient to express the Green's functions as 
\bea\label{greensep}
{\cal G}(x,x')=\int\frac {d\omega}{ 2\pi}\,e^{i\omega(\tau-\tau')}\,G(x^j,x'^j,\omega)~,
\eea where $G(x^j,x'^j,\omega)$ depends on spatial coordinates and $\omega$ only. With the help of the identities (\ref{dets}) and (\ref{comder}), one can easily show that $G$ must satisfy the equation
\bea\non
(\hat\nabla^2-m^2+\omega^2-\epsilon^2\hat\nabla^4)\, G(x^j,x'^j,\omega)=\\
=-q^{-1/2}\delta^{(n)}(x^j-x'^j)~,\label{spaceGreen}
\eea
provided the metric (\ref{Jmetric}) is assumed.

This equation describes the Green's functions associated to a scalar field, which propagates on the spatial sections of the metric (\ref{Jmetric}), as seen by a free-falling observer co-moving with $u^{\mu}$. From the mathematical point of view, we can construct the heat kernel functional $H(s,\Delta x^j)$, formally defined as
\bea
G(\Delta x^j,\omega)=\int_0^{\infty} H(s,\Delta x^j) ds~.
\eea 
Then, $H$ must satisfy the heat equation 
\bea
{\partial H\over \partial s}=-\hat Z H~,
\eea
where $\hat Z$ is an elliptic operator. In our case, $\hat Z=\hat\nabla^2-m^2+\omega^2-\epsilon^2\hat\nabla^4$. 

In the relativistic case, $\epsilon=0$, one can show that the heat kernel has the form \cite{fulling}
\bea\non
H(s,\Delta x^j,\epsilon=0)\sim (4\pi s)^{-n/2}\,e^{-\Delta x^2/4s+s(\omega^2-m^2)}\\
\times\sum_{l=0}^{\infty}\,\hat a_l(x^j,x'^j)s^l~,\label{ansatz}\eea where $n$ is the number of space-like dimensions and the $\hat a's$ are geometric coefficients built on the Riemann tensor associated to $q_{\mu\nu}$, which can be found by a recursion procedure (see \cite{Christensen}). Finally, by integrating over $\omega$, one find the full Green's functions expansion
\bea\non
{\cal G}(x,x')=i\int_0^{\infty} ids(4\pi is)^{-(n+1)/2}\,e^{-\sigma/2is-ism^2}\\
\times\sum_{l=0}^{\infty}\hat a_l(x^j,x'^j)(is)^l~,\label{Grfunc}\eea where we changed $s\rightarrow is$, and set $\sigma={1\over 2}(x^{\mu}-x'^{\mu})^2$. 

A crucial remark is in order here. Despite the above formula looks identical to the usual deWitt-Schwinger expansion, there are major differences. First,  the $\hat a$'s depend on geometrical quantities built on $q_{\mu\nu}$, not $g_{\mu\nu}$. This is due to the fact that the metric (\ref{Jmetric}) is not the most general one, as the space-time under consideration is static with respect to the free-falling time $\tau$. As a consequence, the frequency $\omega$ is not a generic one, but it is the one measured by the free-falling observer only.

To obtain the deWitt-Schwinger expansion in terms of geometrical quantities related to $g_{\mu\nu}$, some extra work is required. As an example, consider the expansion up the the second order only. The only non-trivial coefficient reads \cite{fulling}
\bea
\hat a_1={1\over 6}\hat R~,
\eea where $\hat R$ is the $n$-dimensional Ricci scalar built on $q_{\mu\nu}~$. By contracting the Gauss-Codacci identity \cite{HE}
\bea
\hat R^{\,\alpha}_{~\beta\gamma\delta}=R^{\,\sigma}_{~\mu\nu\rho}q^{\,\alpha}_{~\sigma}q^{\,\mu}_{~\beta}q^{\,\nu}_{~\gamma}q^{\,\rho}_{~\delta}-K^{\alpha}_{~\gamma}K_{\beta\delta}+K^{\alpha}_{~\delta}K_{\beta\gamma}~,
\eea where $K_{\alpha\beta}=q^{\,\gamma}_{~\alpha}\,q^{\,\delta}_{~\beta}\,\nabla_{\delta}\,u_{\gamma}$ is the extrinsic curvature of the spatial sections orthogonal to $u^{\mu}$, and $R^{\,\sigma}_{~\mu\nu\rho}$ is the Riemann tensor constructed with $g_{\mu\nu}$, we can relate $\hat R$ to the curvature of the full space-time, as
\bea
\hat R=R+2R_{\alpha\beta}u^{\alpha}u^{\beta}-(K^{\alpha}_{~\alpha})^2+K^{\alpha\beta}K_{\alpha\beta}~.
\eea Thus, Eq.~(\ref{Grfunc}) contains extra terms with respect to the usual expression \cite{Christensen}
\bea\non
{\cal G}(x,x')=i\int_0^{\infty} ids(4\pi is)^{-(n+1)/2}\,e^{-\sigma/2is-ism^2}\\
\times\sum_{l=0}^{\infty}a_l(x,x')(is)^l~.\eea 

We now turn to the dispersive case. When $\epsilon\neq 0$, the ansatz (\ref{ansatz}) simply does not work, and  the reason will be clear at the end of next section (see also \cite{fulling}). As our $n$-dimensional problem can be seen as the Euclidean continuation of a $n$-dimensional theory with a $\square^2$ term, one can find the heat kernel by the method presented in \cite{Lee}. However, as we will see in the next section, there is an alternative route, which simplifies the calculations and keep track closely of the physics behind the mathematical structure.

\section{Momentum space representation}

In this section, we find a momentum-space representation of the Green's functions, which, in principle, can work for any analytic dispersion relation ${\cal F}(k^2)$. For the moment, we restrict ourself to the quartic case (\ref{MDR}), and discuss further generalizations in the last section. The method that we are going to use is due to Bunch and Parker \cite{BirDav,BP}, and it essentially makes use of a local Taylor's expansion of the metric tensor expressed in Riemann normal coordinates \cite{Christensen, Poisson}. In turn, this leads to a similar expansion of the various differential operators in terms of ordinary partial derivatives and geometrical coefficients.
To begin with, we define the function $\bar G$ such that
\bea
 G(x,x',\omega)=q^{-1/4}(x)\bar G(x,x',\omega)q^{-1/4}(x')~,
 \eea
and write
\bea
q^{-1/2}(x)\delta(x-x')=q^{-1/4}(x)\delta(x-x')q^{-1/4}(x')~.
\eea
These function behaves as a bi-scalars at $x$ and $x'$. For notational convenience, we drop the indices, and $x$ and $x'$ are considered as separate point on the \emph{same spatial slice} $\tau=$ const. Next, we introduce the Riemann normal coordinates $y$ with origin at $x'$. Thus, $q(x')=1$ and the Green's function $\bar G$ satisfies (from now on, the dependence of $G$ from $\omega$ will be understood)
\bea\label{redgreen}
q^{1/4}(\omega^2+m^2-\hat\nabla^2+\epsilon^2\hat\nabla^4)(q^{-1/4}\bar G(y))=\delta(y)~.
\eea  In a neighborhood of $x'$, we can expand the induced metric as \cite{BP}
\bea\non
q_{mn}=\delta_{mn}-{1\over 3}\hat R_{manb}y^ay^b-{1\over 6}\hat R_{manb;p}y^ay^by^p+\\
+\left(-{1\over 20}\hat R_{manb;pq}+{2\over 45}\hat R_{ambl}\hat R^{\,l}_{~pnq}\right)y^ay^by^py^q~,
\eea
from which it follows that
\bea\non
&q&=1-{1\over 3}\hat R_{ab}y^ay^b-{1\over 6}\hat R_{ab;c}y^ay^by^c+\\\non
&+&\left({1\over 18}\hat R_{ab}\hat R_{cd}-{1\over 90}\hat R_{pab}{}^{q}\hat R^{p}_{~cdq}-{1\over 20}\hat R_{ab;cd}\right)y^ay^by^cy^d\,.\\
\eea 
All coefficients are evaluated at $x'$ (i.e. at $y=0$) and contain up to four derivatives of the metric $q_{ij}$. If we write 
\bea\non
&&q^{1/4}\hat\nabla^2(q^{-1/4}\bar G(y))=q^{ij}\partial_i\partial_j\bar G+\partial_i q^{ij}\partial_{j}\bar G+\\\non
&-&\Big[{1\over 16}q^{ij}\partial_i(\ln q)\partial_j(\ln q)+\\
&+&{1\over 4}q^{ij}\partial_i\partial_j(\ln q)+{1\over 4}\partial_iq^{ij}\partial_j(\ln q)\Big]\bar G~,
\eea
we can evaluate all the coefficients and find, up to fourth order,
\bea\non
q^{1/4}\hat\nabla^2(q^{-1/4}\bar G(y))&\simeq& \delta^{ij}\partial_i\partial_j\bar G+{1\over 6}\hat R\bar G+{1\over 6}\hat R_{;j}y^j\bar G+\\
&+&\hat H_{ij}y^iy^j\bar G~,
\eea where
\bea\non
\hat H_{ij}&=&-{1\over 30}\hat R^{p}_{~i}\hat R_{pj}+{1\over 60}\hat R^{p}_{~i}{}^{q}_{~j}\hat R_{pq}+{1\over 60}\hat R^{pql}{}_{i}\hat R_{pqlj}+\\
&+&{3\over 40}\hat R_{;ij}+{1\over 40}\hat R_{ij;p}{}^p~.
\eea The expansion above is also obtained by using the fact that $\bar G$ depends on $y^2$, being rotationally invariant. To find the expansion for the quartic operator, we can proceed by iteration, but first it is convenient to move into momentum space. Hence, we define the local Fourier transform of $\bar G$ as
\bea\label{fourier}
\bar G(y)=\int{d^nk\over (2\pi)^n}\,e^{ik\cdot y}\tilde G(k)~,
\eea
Now, let
\bea
\psi=q^{1/4}\hat\nabla^2(q^{-1/4}\bar G(y))~.
\eea
In momentum space, this corresponds to
\bea\non
\tilde \psi&=&-k^2\tilde G+{1\over 6}\hat R\tilde G+{i\over 3}\hat R_{;j}k^jD\tilde G-2\hat HD\tilde G+\\
&-&4\hat H_{ij}k^ik^jD^2\tilde G~,\label{secondexp}
\eea where we defined the operator $D$ such that
\bea\label{Dop}
\frac{\partial}{\partial k_j}=2k^j\frac{\partial}{\partial k^2}\equiv 2k^jD~.
\eea
In coordinate space, $\hat\nabla^2(q^{-1/4}\bar G(y))=q^{-1/4}\psi$ and 
\bea
q^{1/4}\hat\nabla^4(q^{-1/4}\bar G)=q^{1/4}\hat\nabla^2(q^{-1/4}\psi)~.
\eea As also $q^{-1/4}\psi$ is rotationally invariant, all we need to do is to insert the expansion (\ref{secondexp}) into itself, to find that the Fourier transform of $q^{1/4}\hat\nabla^4(q^{-1/4}\bar G)$ is given by
\bea\non\tilde\psi^{(2)}&=&k^4\tilde G-{k^2\over 3}\hat R\tilde G-{1\over 3}\hat R_jk^j(\tilde G+2k^2D\tilde G)+{1\over 36}\hat R^2\tilde G\\\non
&+&2\hat H(\tilde G+2k^2D\tilde G)+8\hat H_{ij}k^ik^j(k^2D^2\tilde G+D\tilde G)~,\\
\label{psi2}
\eea where $\hat H\equiv \delta^{ij}\hat H_{ij}~$. Note that this iterative procedure can be used to find the expansion, in momentum space, of $\hat\nabla^{2p}$ for any integer $p$.

With these elements, we can expand Eq.~(\ref{redgreen}) in momentum space up to four derivatives of the metric as
\begin{widetext}
\bea\non
&&(k^2+\epsilon^2k^4+m^2-\omega^2)\tilde G-{1\over 6}\hat R\tilde G(1+2\epsilon^2 k^2)-{i\over 3}\hat R_{;j}k^j(D\tilde G+\epsilon^2\tilde G+2\epsilon^2k^2D\tilde G)+{\epsilon^2\over 36}\hat R^2\tilde G+\\\non
&&2\hat H(D\tilde G+\epsilon^2\tilde G+2\epsilon^2k^2D\tilde G)+4\hat H_{ij}k^ik^j(D^2\tilde G+2\epsilon^2k^2D^2\tilde G+2\epsilon^2D\tilde G)=1~.
\eea 
\end{widetext}
At the zeroth order, this equation yields
\bea
\tilde G_0=\frac{1}{k^2+\epsilon^2k^4+m^2-\omega^2}~,
\eea while the following orders can be found by recursion, yielding
\bea
\tilde G_2&=&{1\over 6}\hat R(1+2\epsilon^2 k^2)\tilde G_0^2~,\\
\tilde G_3&=&{1\over 3}\hat R_{;j}k^j\tilde G_0(\tilde G+2k^2D\tilde G_0)~,\\\non
\tilde G_4&=&{1\over 36}\hat R^2(1+2\epsilon^2k^2)^2\tilde G_0^3-{\epsilon^2\over 36}\hat R^2\tilde G_0^2+\\\non
&+&2\hat H\tilde G_0(D\tilde G_0+2\epsilon^2k^2D\tilde G_0+\epsilon^2\tilde G_0)+\\\non
&+&4\hat H_{ij}k^ik^j\tilde G_0(D^2\tilde G_0+2\epsilon^2k^2D^2\tilde G_0+2\epsilon^2D\tilde G_0)~.\\
\eea 
Therefore, as $\tilde G=\tilde G_0+\tilde G_2+\ldots$, we finally have
\bea\non
\tilde G&=&\tilde G_0-{1\over 6}\hat RD\tilde G_0-{i\over 12}\hat R_{;j}\tilde\partial^jD\tilde G_0+\\
&+&\left({1\over 72}\hat R^2-{1\over 3} \hat H\right)D^2\tilde G_0-{1\over 3}\hat H_{ij}\tilde\partial^i\tilde\partial^jD\tilde G_0~,\label{finalexp}
\eea 
where, tilded derivatives are with respect to $k^i$, and where many cancellations occur by using the identities
\bea
D\tilde G_0&=&-(1+2\epsilon^2k^2)\tilde G_0^2~,\\\non\\
(1+2\epsilon^2k^2)D^2\tilde G_0^2&=&-4\epsilon^2D\tilde G_0^2-D^3\tilde G_0~,\\\non\\
\tilde G_0D^2\tilde G_0&=&{1\over 3}D^2\tilde G_0^2-{2\over 3}\epsilon^2\tilde G_0^3~.
\eea
The expansion (\ref{finalexp}) is our main result. 
We conclude this section, by connecting the above expansion with the proper time formalism of deWitt-Schwinger. By using Eqs.~(\ref{greensep}),  (\ref{fourier}), by setting $\omega=k^0$, and by integrating by parts, we find the full $(n+1)$-dimensional Green's function, expressed as
\begin{equation}
{\cal G}(x^{\mu},x'^{\mu})=\int \frac{d^{n+1}k}{(2\pi)^{n+1}}\,e^{ik^{\mu}y_{\mu}}\left[1-f_1D+f_2D^2\right]\tilde G_0~,
\end{equation}
where 
\bea
f_1&=&{1\over 6}\hat R+{1\over 36}\hat R_{;j}y^j-{1\over 3}\hat H_{ij}y^iy^j~,\\\non\\
f_2&=&{1\over 72}\hat R^2-{1\over 3}\hat H~,\eea
are built on $q_{ij}$. These coefficients are formally identical to the ones found by Bunch and Parker in \cite{BP}, but, in that case, they were built on the \emph{full metric} $g_{\mu\nu}$. It is interesting to note that, when, $\epsilon^2=0$ we have 
\bea
D_{\epsilon=0}\equiv \frac{\partial}{\partial m^2}~,
\eea
and the above expression becomes identical to the Bunch and Parker one, except, again, for the geometrical coefficients. As mentioned above, this is due to the fact that we are  working with the  metric (\ref{Jmetric}), which is not the most general one.

At $\omega$ fixed (i.e. on a given spatial slice), we define
\bea\label{srepr}
\tilde G_0=i\int_0^{\infty}ds\, e^{-is(k^2+\epsilon^2k^4+m^2-\omega^2)}~,
\eea and replace into (\ref{finalexp}). By swapping integrals, we find
\bea
\non G(y)&=&i\int_0^{\infty}ds\,e^{-is(m^2-\omega^2)}\Big[1+(is)(1-2\epsilon^2\partial^2)f_1+\\
&+&2(is)^2\epsilon^2f_2+(is)^2(1-2\epsilon^2\partial^2)^2f_2\Big]\, I_{\epsilon}(y,s),\label{dewitt}
\eea where the partial derivatives are with respect to $y$, and
\bea
I_{\epsilon}(y,s)=\int \frac{d^nk}{(2\pi)^n}\,e^{-is(k^2+\epsilon^2k^4)+ik\cdot y}~.
\eea
This integral can be evaluated in form a sum of Hermite polynomials ${\cal H}_{[l]}$ (see the Appendix), namely
\begin{equation}\label{Int}
I_{\epsilon}(y,s)={e^{{iy^2\over 4s}}\over (4is\pi)^{n/2}}\,\sum_{\lambda=0}^{\infty}{1\over \lambda !}\left(\frac{i\epsilon^2}{16s}\right)^{\lambda}{\cal H}_{[4\lambda]}\left(\frac{\vec{y}}{\sqrt{4is}}\right).
\end{equation}
It is clear that the deWitt-Schwinger expansion in proper time (\ref{dewitt}) becomes a very complicate sum of Hermite polynomials and their derivatives, which does not appear to converge to any known function. In the Lorentz invariant case instead, the sum is trivial and the integral over $s$ converges to Hankel functions of second kind \cite{dewitt,Christensen}.  It is now clear why the heat kernel ansatz (\ref{ansatz}) cannot work.

\section{Discussion}

\noindent In this work we present our first results concerning quantum field theory on curved backgrounds, with modified dispersion relations. In particular, we carefully analyze the Klein-Gordon equation associated to a scalar field propagating on a ultra-static space-time, as a first step towards the physically relevant case of stationary metrics. 

In the simplest case, superluminal propagation of high frequency modes can be achieved by adding a quartic spatial derivative to the Klein-Gordon operator. An important side-effect of these higher order operators is the breaking of Lorentz invariance. On physical grounds, this leads to important modifications to the Green's functions, even in flat space. As an example, we considered the 2-dimensional flat space, and we proved that the two-point function becomes finite in the coincidence limit. 

Motivated by this, and also by the results achieved in the context of Cosmology, we decided to consider the problem in a non-homogeneous curved space. 
The problem becomes much more difficult, as in the Klein-Gordon equation we now have higher-derivative spatial operators, in contrast to the cosmological case, which prevent the usual deWitt-Schwinger expansion of the Green's functions from working properly. We turned around this problem by assuming the existence of a preferred frame, encoded by the unit time-like vector $u^{\mu}$, associated to the free-falling observer. In this way, despite the breaking of local Lorentz invariance, we still have general covariance and, above all, rotational invariance over slices of constant time. 
In the case of a ultra-static space-time, this fact allows for a dimensional reduction of the Klein-Gordon equation, so that one can consider unambiguously the Green's functions at a fixed frequency, as measured by the free-falling observer.  As a result, one deals with elliptic operators at constant time. Despite this simplification, the usual heat-kernel ansatz does not work yet because of the higher derivative operators, responsible for the modified dispersion relation. Therefore, it turned out to be much easier to work in momentum-space, following the method of Bunch and Parker. Thus, we obtained an expansion of the Green's functions up to four derivatives of the metric in momentum space. We also showed that the proper-time formulation of our expansion leads to a very complicate expression, which might render very difficult to renormalize the theory in coordinate space.

We wish to conclude with a remark. In this work, we displayed our method for the simplest case, namely a $k^4$ term in the dispersion relation. However, it looks like quite easy to include higher order derivatives. In order to do that, one should compute terms like $\tilde \psi^{(2p)}$, for $p$ integer, by recursion and by exploiting the $O(n)$ invariance of the Green's functions, in the same way as we calculated Eq.~(\ref{psi2}). In the case $p=3$, we obtain few more terms which can be absorbed by a redefinition of the operator $D$. Therefore, the formal expression (\ref{finalexp}) does not change. A generic dispersion relation of the form ${\cal F}(k^2)$, such that ${\cal F}$ is analytic, can always be expanded as a power series in $k^2$. Therefore, it is reasonable to expect that Eq.~(\ref{finalexp}) does not change, provided $D$ is adequately defined. We hope to prove this conjecture in a future work.

\vspace{1cm}
\noindent {\large\bf{Acknowledgements}} 

\noindent I wish to thank R.~Balbinot, A.~Fabbri, S.~Fagnocchi, R.~Parentani, P.~Anderson, A.~Ottewill, M.~Casals, and A.~Ambach for helpful discussions and suggestions. I would like to thank also the Departamento de Fisica Teorica, Valencia U., for hospitality during part of this work.

\appendix
\section{Evaluation of $I_{\epsilon}(y,s)$ }\setcounter{equation}{0}

\noindent To evaluate $I_{\epsilon}(y,s)$, we first write it as
\begin{equation}
I_{\epsilon}(y,s)=\int \frac{d^nk}{(2\pi)^n}\,e^{-isk^2+iky}\,e^{-\gamma k^4}~, \quad \gamma=is\epsilon^2.
\end{equation}
By expanding the second exponential, we can write
\begin{equation}
I_{\epsilon}(y,s)=\sum_{\lambda}^{\infty}\frac{(-\gamma)^{\lambda}}{\lambda !}\left(\frac{\partial}{\partial \vec{y}}\right)^{4\lambda}\int \frac{d^nk}{(2\pi)^n}\,e^{-isk^2+iky}.
\end{equation}
The integral is now a product of $n$ Gaussian integrals, hence
\begin{equation}
I_{\epsilon}(y,s)=(4is\pi)^{-n/2}\sum_{\lambda}^{\infty}\frac{(-\gamma)^{\lambda}}{\lambda !}\left(\frac{\partial}{\partial \vec{y}}\right)^{4\lambda}\,\exp{\left(iy^2\over 4s\right)}~.
\end{equation} By recalling the definition of Hermite polynomials
\bea
{\cal H}_{[l]}=(-1)^l\,e^{\,x^2}\frac{\partial^l}{\partial x^l}\,e^{-x^2}~,
\eea and by changing variable, we finally find Eq.~(\ref{Int}).

\end{document}